\documentclass[aps, twocolumn, showpacs, preprintnumbers, amsmath, amssymb, prb]{revtex4-2}
\usepackage{amsfonts}
\usepackage{mathrsfs}
\usepackage{color}
\usepackage[svgnames]{xcolor}   % ?? SVG ?????
\usepackage{natbib}
\usepackage{graphicx}
\usepackage{bm}% bold maths
\usepackage{amssymb}
\usepackage{epstopdf}
\usepackage{array}
\usepackage[colorlinks=true, letterpaper=true, pdfstartview=FitV, linkcolor=blue, citecolor=blue, urlcolor=blue]{hyperref}
\usepackage{float}
\usepackage[figuresright]{rotating}
\usepackage{orcidlink}%作者加上orcid链接

\makeatletter

\newcommand{\Rmnum}[1]{\expandafter\@slowromancap\romannumeral #1@}
\makeatother
\providecommand{\U}[1]{\protect\rule{.1in}{.1in}}
\newcommand{\CVB}[0]{\color{black}}
\newcommand{\CIVB}[0]{\color{black}}
\newcommand{\CVA}[0]{\color{black}}
\newcommand{\CIVA}[0]{\color{black}}
\begin{document}

\title{Anomalous \CVA Transverse \CIVA Thomson Effect }

%\author{Zhi-Fan Zhang$^{1}$\,\orcidlink{0000-0002-7063-4020}} %作者加上orcid链接

\author{Ying-Fei Zhang$^{1}$\,%\orcidlink{0009-0003-1463-1655}
}
%\altaffiliation{These authors contributed equally to this work.}
\author{Zhi-Fan Zhang$^{2}$ \,%\orcidlink{0000-0002-7063-4020}
}
%\altaffiliation{These authors contributed equally to this work.}

\author{Zhen-Gang Zhu$^{1,3}$\,%\orcidlink{0000-0002-2837-6072}
}
\email{zgzhu@ucas.ac.cn}

\author{Gang Su$^{4,5,1}$\,%\orcidlink{0000-0002-8149-4342}
}
\email{sugang@itp.ac.cn; gsu@ucas.ac.cn}

\affiliation{
$^{1}$School of Physical Sciences, University of Chinese Academy of Sciences, Beijing 100049, China.\\	
$^{2}$Interdisciplinary Center for Theoretical Physics and Information Sciences (ICTPIS), Fudan University, Shanghai 200433, China. \\
$^{3}$School of Electronic, Electrical and Communication Engineering, University of Chinese Academy of Sciences, Beijing 100049, China.\\
$^{4}$Institute of Theoretical Physics, Chinese Academy of Sciences, Beijing 100190, China.\\
$^{5}$ Kavli Institute for Theoretical Sciences, University of Chinese Academy of Sciences, Beijing 100190, China.
}

\date{\today}

\begin{abstract}

 We formulate the anomalous transverse Thomson effect (ATTE), the long-missing Thomson counterpart of the anomalous Hall and anomalous Nernst effects. Its coefficient is not unique but depends on the electrical boundary condition: $\tau_{\rm ATTE}^{(j)}$, defined for a spatially uniform imposed current density, and $\tau_{\rm ATTE}^{(E)}$, defined for equipotential electrodes. Both are fully determined by the total anomalous Nernst coefficient (ANC) and its temperature dependence, while $\tau_{\rm ATTE}^{(E)}$ acquires an additional contribution from the temperature dependence of the longitudinal resistivity. This relation is model independent within local linear response and applies to the total ANC regardless of whether its microscopic origin is intrinsic or extrinsic. Using a massive Dirac model for Fe$_3$Sn$_2$, we show that Berry curvature near the gapped Dirac cones enhances the ANC-related contributions to $\tau_{\rm ATTE}^{(E)}$. In the low-temperature limit, both $\tau_{\rm ATTE}^{(j)}/S_{xy}^{\rm A}$ and $\tau_{\rm ATTE}^{(E)}/S_{xy}^{\rm A}$ approach a value of three. We further extract $\tau_{\rm ATTE}^{(E)}$ from reported experimental data for CoS\(_2\), Co\(_3\)Sn\(_2\)S\(_2\) and CeCrGe\(_3\); for CeCrGe$_3$ it reaches approximately -20 $\mu V/K$, and we estimate the maximum achievable cold-side heat current density to assess its cooling potential. Because the ATTE can be evaluated directly from existing ANC and longitudinal resistivity data, it is readily testable without additional equipment or measurements.
\end{abstract}
\maketitle

\begin{figure*}[tb] %
\centering
\includegraphics[width=0.9\textwidth]{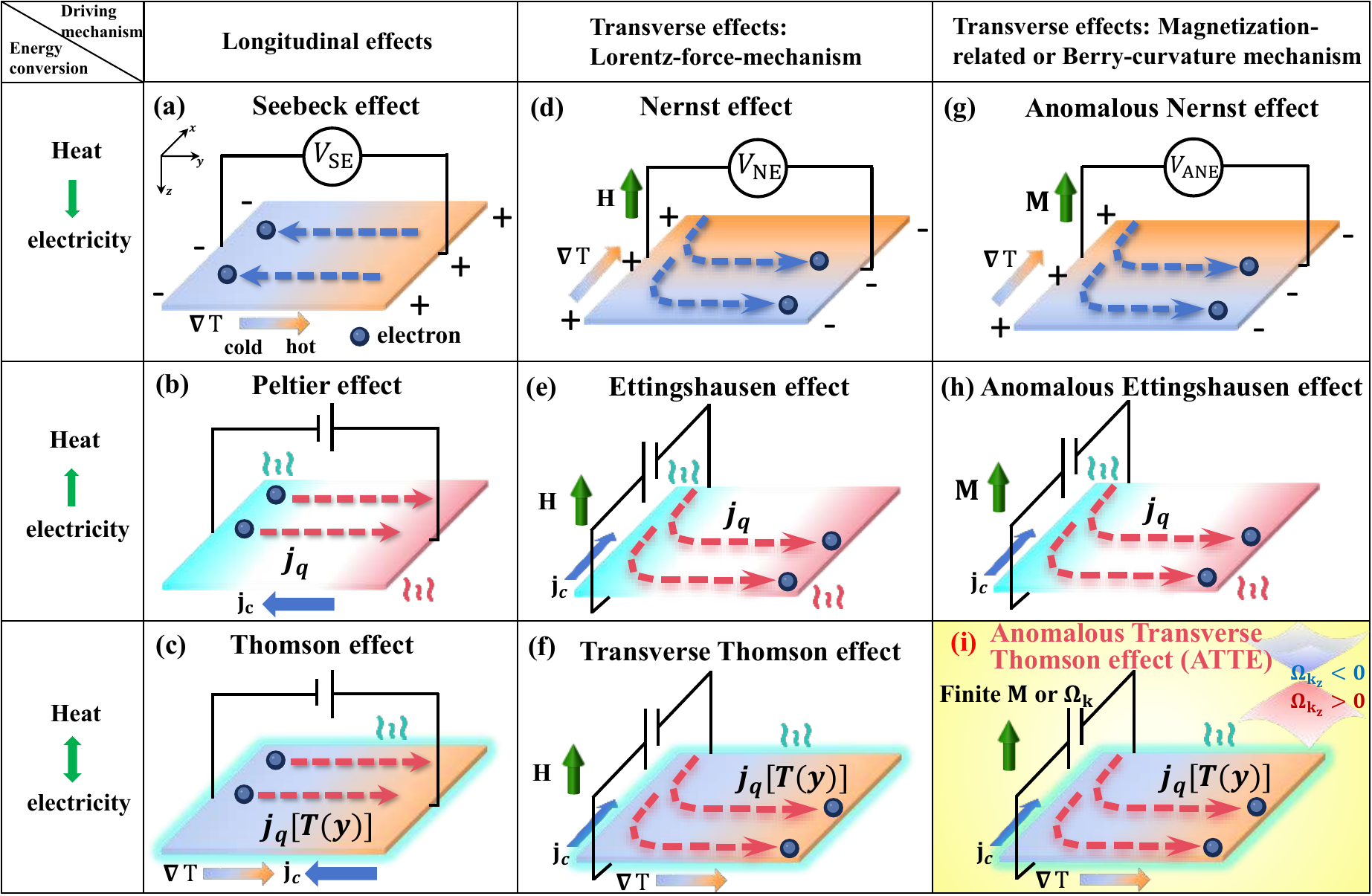} %
\caption{Schematics of thermoelectric, thermomagnetic \CVB and anomalous thermoelectric \CIVB effects. (a-c) \CVB The longitudinal thermoelectric effects driven by generalized thermodynamic forces, including the \CIVB Seebeck, Peltier, and Thomson effects. (d-f) \CVB The transverse thermomagnetic effects driven by the Lorentz force under a magnetic field $\mathbf{H}$, namely the \CIVB Nernst, Ettingshausen, and transverse Thomson effects. (g-i) Anomalous counterparts driven by magnetization $\mathbf{M}$ \CVB or Berry curvature of energy bands,  including the anomalous Nernst, anomalous Ettingshausen, and anomalous \CVA transverse \CIVA Thomson effects. \CIVB Here $\boldsymbol{\nabla} T$ is the applied temperature gradient, $\mathbf j_c$ is the charge current, and $\mathbf j_q$ is the heat current.	}
\label{Fig.1}
\end{figure*}

%\section{INTRODUCTION}
%\CV 
\textit{Introduction.-} %\CIV
Thermoelectrics \cite{Blatt1978,Anatychuk2024,Wang2025,Dai2025} represents a crucial area of condensed matter physics, encompassing phenomena such as Seebeck \cite{Ohta2007,Maekawa2004,Behnia2016}, Nernst \cite{Ettingshausen1886,Behnia2009,Behnia2016,Ramos2014,Weiler2012}, Peltier \cite{DiSalvo1999,Gurevich2005,Cui2017}, and Ettingshausen \cite{Paranjape1960,Adachi2025,Delves1962} effects.
As in Fig. \ref{Fig.1}(a), the Seebeck effect manifests as a longitudinal electric voltage output in response to a longitudinal temperature gradient.
In contrast, the Nernst effect [Fig. \ref{Fig.1}(d)] generates a transverse electric voltage under the same temperature gradient but in the presence of a perpendicular magnetic field $\mathbf{H}$.
Both effects convert thermal energy into electricity.
The Peltier effect [Fig. \ref{Fig.1}(b)] describes the absorption or release of heat at the junction of two different materials driven by an electric current $\mathbf{j}_{c}$, while the Ettingshausen effect [Fig. \ref{Fig.1}(e)] serves as its transverse counterpart. Both are recognized as fundamental mechanisms for solid-state cooling \cite{Mao2020,Razeghi2025}. \CVA
Although less efficient than conventional vapor-compression systems, thermoelectric devices provide compact, solid-state, and reliable cooling without moving parts or refrigerants, enabling applications from automotive thermal management to aerospace systems \cite{Zhao2014,Sofrata1984,Chinguwa2018,Bell2008,Snyder2008}. This broad utility motivates the search for new mechanisms to enhance thermoelectric heat management.

%Although currently less efficient than conventional vapor-compression systems, thermoelectric devices are indispensable for specialized applications due to their compact and solid-state reliability. Operating entirely without moving parts or refrigerants, they are utilized in fields ranging from precise automotive thermal management to deep-space radioisotope thermoelectric generators \cite{Zhao2014,Sofrata1984,Chinguwa2018,Bell2008,Snyder2008}. Such established utility underscores the value of exploring novel fundamental heat-management mechanisms.
\CIVA

In particular, the third fundamental thermoelectric phenomenon, Thomson effect (TE) \cite{Thomson1857,Thomson1857a,Lifsic1993,Young1924,Maekawa2004}, was predicted and verified by Lord Kelvin to complement the Seebeck and Peltier effects.
As illustrated in Fig. \ref{Fig.1}(c), when $\mathbf{j}_c$ is applied and accompanied by $ \boldsymbol{\nabla} T$, the Peltier thermal response \CVB$\mathbf{j}_q[T(y)]$ \CIVB transported by charge carriers evolves as they traverse the varying temperature profile.
Consequently, the carriers must adjust their associated heat content to remain consistent with the local Peltier coefficient. This energy exchange results in a continuous, reversible heat absorption or release throughout the material bulk-occurring alongside irreversible Joule heating-with a sign determined by the relative directions of $\mathbf{j}_c$ and $ \boldsymbol{\nabla} T$.
The volumetric heat production rate is given by $\dot{q} = -\tau_{\text{TE}} \mathbf{j}_c \cdot \boldsymbol{\nabla} T$ where the Thomson coefficient $\tau_{\text{TE}}$ is linked to Seebeck coefficient $S_{\text{0}}$ via the second Kelvin relation, $\tau_{\text{TE}} = T(\text dS_{\text{0}}/\text dT)$ \cite{Apertet2016,Gong2019,Zevalkink2018}.
%
%This relationship implies that the Thomson effect only manifests when $S_{\text{SE}}$ exhibits temperature dependence.
%
By enabling distributed heat management rather than the junction-localized cooling characteristic of the Peltier effect, the Thomson effect serves as a vital mechanism for enhancing the performance of thermoelectric cooling devices \cite{Snyder2012,Giaretto2020,Chiba2023,Zebarjadi2023}. This potential for superior efficiency has led to a recent resurgence of interest in the field.
%%%%%%%%%%%%%%%%%%%%%%%%%%%%%%%%%%%%%%%%%%%%%%%%%%%%%%%%%%%%%%%%%%%%%%%%%%%%%%%%%%%%%%%%%%%%%%%%%%%%%%%%%%%%%%%%%%%%%%%%%%%%%%%%%%%%%%%%%%%%%%%%%%%%%%%%%%%%%%%%%%%%

Recently, Uchida \textit{et al.} \cite{Uchida2020} observed the magneto-Thomson effect in Bi$_{88}$Sb$_{12}$ alloys, demonstrating the ability to control the Thomson coefficient through a magnetic field in a non-magnetic material under $\mathbf{j}_c || \boldsymbol{\nabla} T$.
%
%\CV
Subsequently,  the anisotropic magneto-Thomson effect  \cite{Modak2023} is demonstrated in ferromagnetic NiPt, where the Thomson response depends on the magnetization orientation $\mathbf{M}$ in the same geometry $\mathbf{j}_c || \boldsymbol{\nabla} T$. %\CIV
Moreover, experimental findings confirm the presence of the transverse Thomson effect \CVB (TTE) \CIVB \cite{Takahagi2025}.
This effect was observed under specific conditions, i.e. $\mathbf{H}\perp\mathbf{j}_c\perp\boldsymbol{\nabla} T$, as shown in Fig. \ref{Fig.1}(f).
This transverse thermoelectric effect is highly advantageous for efficiently harvesting thermal energy across large surface areas, while also offering a simplified design and reduced manufacturing complexity and cost \cite{Adachi2025,Takahagi2025,Uchida2022}.

%%%%%%%%%%%%%%%%%%%%%%%%%%%%%%%%%%%%%%%%%%%%%%%%%%%%%%%%%%%%%%%%%%%%%%%%%%%%%%%%%%%%%%%%%%%%%%%%%%%%%%%%%%%%%%%%%%%%%%%%%%%%%%%%%%%%%%%%%%%%%%%%%%%%%%%%%%%%%%%%%%%%
%%%%%%%%%%%%%%%%%%%%%%%%%%%%%%%%%%%%%%%%%%%%%%%%%%%%%%%%%%%%%%%%%%%%%%%%%%%%%%%%%%%%%%%%%%%%%%%%%%%%%%%%%%%%%%%%%%%%%%%%%%%%%%%%%%%%%%%%%%%%%%%%%%%%%%%%%%%%%%%%%%

%\CV Here we identify and formulate the anomalous Thomson effect (ATE), the anomalous counterpart of the transverse Thomson effect. The recently observed transverse Thomson effect is governed by the ordinary Nernst coefficient \(S_{xy}\), which is induced by the Lorentz force and is odd in the applied magnetic field in the ordinary weak-field regime. By contrast, the ATE is governed by the anomalous Nernst coefficient \(S^A_{xy}\), which is allowed by broken time-reversal symmetry and anomalous transverse transport. Therefore, after the magnetic or antiferromagnetic domain state is set, the ATE can operate without an external magnetic field. This distinction is analogous to that between the ordinary Hall/Nernst effects and their anomalous counterparts. Microscopically, \(S^A_{xy}\) may arise from intrinsic Berry curvature, extrinsic skew or side-jump mechanisms, or other anomalous transport channels; the model calculation below focuses on the intrinsic Berry-curvature contribution. \CIV

We \CVB formulate \CIVB the anomalous \CVA transverse \CIVA Thomson effect \CVA(ATTE)\CIVA, \CVB which is the Thomson counterpart of \CIVB the anomalous Hall effect (AHE) \cite{Nagaosa2010,Xiao2010} and anomalous Nernst effect (ANE) \CVA [Fig.\ref{Fig.1}(g)] \CIVA \cite{Xiao2006}.
By using the ferromagnetic Dirac system Fe$_3$Sn$_2$ as a case study for specific predictions, \CVB the model calculation focuses on the intrinsic Berry curvature $\mathbf\Omega_{k}$.
As a topological property of Bloch states, $\mathbf\Omega_{k}$ effectively behaves as a magnetic field in momentum space, leading to transverse transport.
 We further introduce the maximum cold-side heat-current density as a cooling capacity metric for the  ATTE within our theoretical framework.  \CVA Our principal contribution is formulating this previously unrecognized bulk thermodynamic response; these metrics serve to connect the ATTE to conventional parameters rather than substituting for a practical refrigerator demonstration. \CIVA
%%%%%%%%%%%%%%%%%%%%%%%%%%%%%%%%%%%%%%%%%%%%%%%%%%%%%%%%%%%%%%%%%%%%%%%%%%%%%%%%%%%%%%%%%%%%%%%%%%%%%%%%%%%%%%%%%%%%%%%%%%

 \textit{Formalism of the Anomalous \CVA Transverse \CIVA Thomson Effect (\CVA ATTE\CIVA).-}
In certain  magnetic materials with spontaneous magnetization $\mathbf{M}$ and a finite $\mathbf\Omega_{k}$, the \CVA ATTE \CIVA arises when $\mathbf{j}_c$ and $\boldsymbol{\nabla} T$ are applied mutually perpendicularly, as schematically shown in Fig. \ref{Fig.1}(i).
Specifically, the \CVA ATTE \CIVA originates from \CVB the temperature evolution of the anomalous Ettingshausen heat current\CIVB, where $\mathbf{j}_c$ drives a transverse thermal current response $\mathbf{j}_{q}$ along the direction of accumulated temperature gradient [Fig. \ref{Fig.1}(h)]. \CVB A net magnetization is sufficient but not necessary; the same phenomenology applies to systems such as noncollinear antiferromagnets \cite{Ikhlas2017} whenever symmetry allows a nonzero \(S^{\text A}_{xy}\). \CIVB
%
%Due to the temperature dependence of the transport coefficients, this thermal current becomes spatially non-uniform within the temperature distribution. Consequently, as charge carriers traverse the temperature gradient, they continuously absorb or release heat to re-equilibrate with the local energy states at new positions, manifesting as a reversible volumetric heating or cooling within the bulk material.
%
%%%%%%%%%%%%%%%%%%%%%%%%%%%%%%%%%%%%%%%%%%%%%%%%%%%%%%%%%%%%%%%%%%%%%%%%%%%%%%%%%%%%%%%%%%%%%%%%%%%%%%%%%%%%%%%%%%%%%%%%%%%%%%%%
%
Combining Ohm's law \cite{Kittel2013}, Fourier's law \cite{Ashcroft2012}, thermoelectric effects relations \cite{Ziman1972,Behnia2016,Behnia2009}  and the Onsager reciprocity relations \cite{Wolfe1963,Gan2021}, we get \cite{Landau1961}
\begin{align}\label{eq:matric}
	\begin{pmatrix}
		\mathbf{E}\\
		\mathbf{j}_{\text{q}}
	\end{pmatrix} =\begin{pmatrix}
		\hat\rho  & \hat S\\
		\hat S T & -\hat\kappa'
	\end{pmatrix}\begin{pmatrix}
		\mathbf{j}_{\text{c}}\\
		\boldsymbol{\nabla} T
	\end{pmatrix},
\end{align}
where $\mathbf{E}$ is the electrical field, \CVB the tensors \(\hat\rho\), \(\hat S\), and \(\hat\kappa'\) denote the electrical resistivity, thermoelectric coefficient, and effective thermal conductivity. We consider an in-plane isotropic system, so that
\(\rho_{xx}=\rho_{yy}\equiv\rho_0\), \(S_{xx}=S_{yy}\equiv S_0\), and \(\kappa'_{xx}=\kappa'_{yy}\equiv\kappa'_0\), whereas
\(S^{\text A}_{xy}=-S^{\text A}_{yx}\) and \(\kappa'_{xy}=-\kappa'_{yx}\).
Thus \(S_0\) is the Seebeck coefficient and \(S^{\text A}_{xy}\) is the anomalous Nernst coefficient \CVA (ANC)\CIVA. The tensor \(\kappa'\) is given by
\(\hat\kappa'=\hat\kappa-\hat\sigma (\hat S^{\text A})^2T\), with \(\hat\kappa\) the thermal conductivity. \CIVB

%Consider $\mathbf j_c$ along $x$ direction and $\nabla_y T$, the volumetric heat generation rate  \cite{Zebarjadi2023,Gong2019,Chiba2023,Sun2020,Apertet2016} can be expressed as $\dot{q}_{\text{tot}}  =  E_{x} j_{\text{c} ,x} - \CVB \nabla_{y} \CIVB j_{\text{q} ,y}$, and we have \cite{sm}
%\begin{align}
%\label{eq:dot q_ATTE}
%	\dot{q}_{\text{tot}} & =E_{x} j_{\text{c} ,x} -\text{div}  j_{\text{q} ,y}\nonumber\\
%	& =\rho _{xx} j_{\text{c} ,x}^{2} +\nabla _{y} (\kappa '_{xx} \nabla _{y} T)+\left(  2S_{xy}^{A} +T\frac{\mathrm{d} S_{xy}^{A}}{\mathrm{d} T}\right) j_{\text{c} ,x} \nabla %_{y} T,
%\end{align}
%\begin{equation}\label{eq:dot q_ATTE-1}	\dot{q}_{\text{tot}}  %& = & E_{x} j_{\text{c} ,x} -\text{div}  j_{\text{q} ,y}  \notag\\	 =\CVB \rho _{0} \CIVB j_{\text{c} ,x}^{2} +\nabla _{y} ( \kappa '_{0}  \nabla _{y} T)+ \CVA\tau_{\text{ATTE}}\CIVA j_{\text{c} ,x} \nabla _{y} T,\end{equation}

\CVA For $\mathbf j_c=j_{c,x}(y)\hat{\mathbf x}$ and $\boldsymbol\nabla T=(\nabla_yT)\hat{\mathbf y}$, the volumetric heat generation rate  \cite{Zebarjadi2023,Gong2019,Chiba2023,Sun2020,Apertet2016} can be expressed as $\dot{q}_{\text{tot}}  =  E_{x} j_{\text{c} ,x} -  \nabla_{y}  j_{\text{q} ,y}$, and we have \cite{sm}
 \begin{equation}
	\label{eq:dot q_ATTE-1}
	\dot{q}_{\text{tot}} 	=\rho _{0} j_{\text{c} ,x}^{2} +\nabla _{y} (\kappa '_{0}\nabla _{y} T)
	+\tau_{\text{ATTE}}^{(j)}j_{\text{c} ,x}\nabla _{y} T
	+T S_{xy}^{\text A}\nabla_y j_{c,x},
\end{equation} where the first two terms are Joule heating and heat conduction. The last two terms constitute the reversible anomalous transverse contribution, i.e., ATTE.

%Electrical boundary conditions should be specified before the last term of Eq.~(\ref{eq:dot q_ATTE-1}) is absorbed into a coefficient. For a spatially uniform imposed current density, it is

Before the last term of Eq. (\ref{eq:dot q_ATTE-1}) can be absorbed into a Thomson-type coefficient, the electrical boundary condition must be specified, because different boundary conditions define different coefficients. For a spatially uniform imposed current density, the ATTE term reads
\begin{align}
	\label{eq:q_ATTE}
	\dot{q}_{\text{ATTE}} = -\tau _{\text{ATTE}}^{(j)}\left(\mathbf{m} \times \mathbf{j}_c \right) \cdot \boldsymbol{\nabla} T.
\end{align} \CIVA
%
%where the first term  accounts for Joule heating, the second term corresponds to heat conduction, and the third term proportional to $j_{c,x}\nabla_yT$ arises from the introduced \CVA ATTE \CIVA in this work, which is defined as
%\begin{align}\label{eq:q_ATTE}	\dot{q}_{\text{ATTE}}  = -\CVA\tau _{\text{ATTE}}\CIVA\left(\mathbf{m} \times \mathbf{j}_c \right) \cdot \boldsymbol{\nabla} T.\end{align}
The \CVA ATTE  is finite when \((\mathbf m\times\mathbf j_c)\cdot\boldsymbol\nabla T\neq0\) with $\mathbf{m}={\mathbf{M}}/{|\mathbf{M} |}$. \CIVA The cooling or heating condition is governed by the combined sign of $\CVA\tau^{(j)}_{\rm ATTE}\CIVA$ and the geometrical factor
$\left(\mathbf{m}\times\mathbf j_c\right)\cdot \boldsymbol\nabla T$. 
After a derivation \cite{sm}, the ATTE coefficient is obtained as
\begin{eqnarray}
	 \CVA\tau^{(j)}_{\text{ATTE}}\CIVA &=& T \text{d} S_{xy}^{\text{A}}/\text{d} T +2S_{xy}^{\text{A}}, \label{eq:tau_ATTE} \\
%\end{align}
%where
%\begin{align}
%
S^{\text{A}}_{xy} &=& {(\sigma_{xx}\alpha^{\text{A}}_{xy}-\sigma^{\text{A}}_{xy}\alpha_{xx})}/{(\sigma_{xx}^2+(\sigma^{\text{A}}_{xy})^{2})}. \label{eq:S_xy}
\end{eqnarray}
$S^{\text{A}}_{xy}$ is the ANC \cite{Ikhlas2017,Li2023,Ding2019,Chen2025}, determined by longitudinal electrical conductivity $\sigma _{xx}$, longitudinal thermoelectric coefficients $\alpha_{xx}$, anomalous Hall conductivity $\sigma^{\text{A}} _{xy}$ \cite{Xiao2010,Nagaosa2010}, and anomalous thermoelectric coefficients $\alpha^{\text{A}} _{xy}$ \cite{Xiao2006}. %, as these are essential components in determining the overall response.
%%%%%%%%%%%%%%%%%%%%%%%%%%%%%%%%%%%%%%%%%%%%%%%%%%%%%%%%%%%%%%%%%%%%%%%%%%%%%%%%%%%%%%%%%%%%%%%%%%%%%%%%%%%%%%%%%%%%%%%%%%%%%%%%%%%%%%%%%%%%%%%%%%%%
%
\CVB Eq. (\ref{eq:tau_ATTE}) follows from local energy balance and Onsager thermoelectric relations. Thus \(S^{\text A}_{xy}\) denotes the total \CVA ANC\CIVA, regardless of whether it originates from Berry curvature, skew scattering, side jump, or other anomalous mechanisms. \CIVB %which enter the ATE only through \(S^A_{xy}\)
%Eq. (\ref{eq:tau_ATTE}) represents a general relation independent of the specific model and scattering mechanism.
%
%The ATC is determined not only by ANC $S^{\text{A}}_{xy}$  but also by its temperature derivative.
%
%Thus, the ATE is a distinct phenomenon from ANE.
%
%It is reasonable to expect that the ATE may remain prominent due to the temperature derivative even when $S_{xy}^{\text{A}}$ is small.
%
%\CVA These anomalous mechanisms enter the ATTC through the total $S_{xy}^\text A$, while the longitudinal charge transport physics additionally enters through the temperature dependence of $\sigma_{xx}$. The ATTC contains three contributions.   \CIVA
%
The first term in Eq.~(\ref{eq:tau_ATTE}), $T\text dS_{xy}^\text A/\text dT$ describes the temperature evolution of the ANC. The second term 2$S_{xy}^{\text{A}}$ originates from local energy balance. One contribution to $S_{xy}^{\text{A}}$ arises from electrical work, i.e., $\mathbf{E}\cdot\mathbf{j}_c$;
while another $S_{xy}^{\text{A}}$ is due to the divergence of the %Ettingshausen 
heat current $\mathbf{j}_q$.
 The 2$S_{xy}^{\text{A}}$ term is unique to transverse effects and is absent in the longitudinal TE \cite{Landau1961,Takahagi2025}.

 \CVA For equipotential longitudinal electrodes, the longitudinal electric field is spatially uniform along the transverse direction, $\nabla_yE_x= 0$, retaining the bilinear term ($j_{c,x}\nabla_yT$) in Eq. (\ref{eq:dot q_ATTE-1}) yields
  \begin{equation}
 	\tau_{\rm ATTE}^{(E)}=\tau_{\rm ATTE}^{(j)}-T S_{xy}^{\rm A}{\text d\ln\rho_{xx}}/{\text dT}.
 	\label{eq:tau_ATTE_E}
 \end{equation}
%The term $TS_{xy}^{\text A}\text d\ln\rho_{xx}/\text dT$, originates from the spatial redistribution of the current density $j_{c,x}$ caused by the temperature dependence of $\rho_{xx}(T)$. More explicitly, $\rho_{xx}[T(y)]$ gives $ \nabla_yj_{c,x}=-j_{c,x}(\text d\ln \rho_{xx}/\text dT)\nabla_yT$, and this coefficient $\tau_{\rm ATTE}^{(E)}$ reduces to $\tau_{\rm ATTE}^{(j)}$ for $T$-independent $\rho_{xx}$ or a locally uniform imposed current. 
The term $TS_{xy}^{\text A}\text d\ln\rho_{xx}/\text dT$ originates from the spatial redistribution of the longitudinal current density $j_{c,x}$ induced by the temperature dependence of $\rho_{xx}$. Indeed, since $\rho_{xx}$ depends on position through $T(y)$, a transverse temperature gradient produces $ \nabla_yj_{c,x}=-j_{c,x}(\text d\ln \rho_{xx}/\text dT)\nabla_yT$ at this order, which is precisely the redistribution that distinguishes $\tau_{\rm ATTE}^{(E)}$ from $\tau_{\rm ATTE}^{(j)}$. Accordingly, $\tau_{\rm ATTE}^{(E)}$ reduces to $\tau_{\rm ATTE}^{(j)}$ when $\rho_{xx}$ is temperature independent, or when $\nabla_yT=0$ so that the redistribution vanishes.

\CIVA

 Notably, Eq.~(\ref{eq:tau_ATTE}) shares the same algebraic structure as the TTE \cite{Takahagi2025}. The underlying mechanisms fundamentally differ, where the TTE stems from the Lorentz-force-induced ordinary Nernst coefficient and vanishes in the absence of an external magnetic field, whereas the ATTE is governed by $S^{\text A}_{xy}$ and can persist at zero field \CVA after the (antiferro)magnetic or noncollinear domain state is established. Eq.~(\ref{eq:tau_ATTE_E}) additionally shows how a specified electrode geometry can modify the ATTE through current redistribution. \CIVA It is thus not a simple extension of the TTE, but a distinct bulk heat-generation/absorption response inherent to anomalous systems.

%Notably, \CVA the first two terms of \CIVA  Eq. (\ref{eq:tau_ATTE}) share the same algebraic structure as the TTE \cite{Takahagi2025}, \CVA In addition, Eq. (4) retains a conductivity-gradient contribution, $S_{xy}^{\text A}T\text d\ln\sigma_{xx}/\text d T$. \CIVA  The underlying mechanisms fundamentally differ, where the TTE stems from the Lorentz-force-induced ordinary Nernst coefficient and vanishes in the absence of an external magnetic field, whereas Eq. (\ref{eq:tau_ATTE}) \CVA is governed by the ANC $S^{A}_{xy}$ that can persist at zero field, with its magnitude additionally modulated by the temperature dependence of the $\sigma_{xx}$.  \CIVA Therefore, once the (antiferro)magnetic domain configuration is established, the \CVA ATTE \CIVA can operate without involving any external $\mathbf{H}$ field. It is thus not a simple extension of the TTE, but a distinct bulk heat-generation/absorption response inherent to anomalous systems.

\begin{figure}[tb]
	\centering
	\includegraphics[width=1\linewidth]{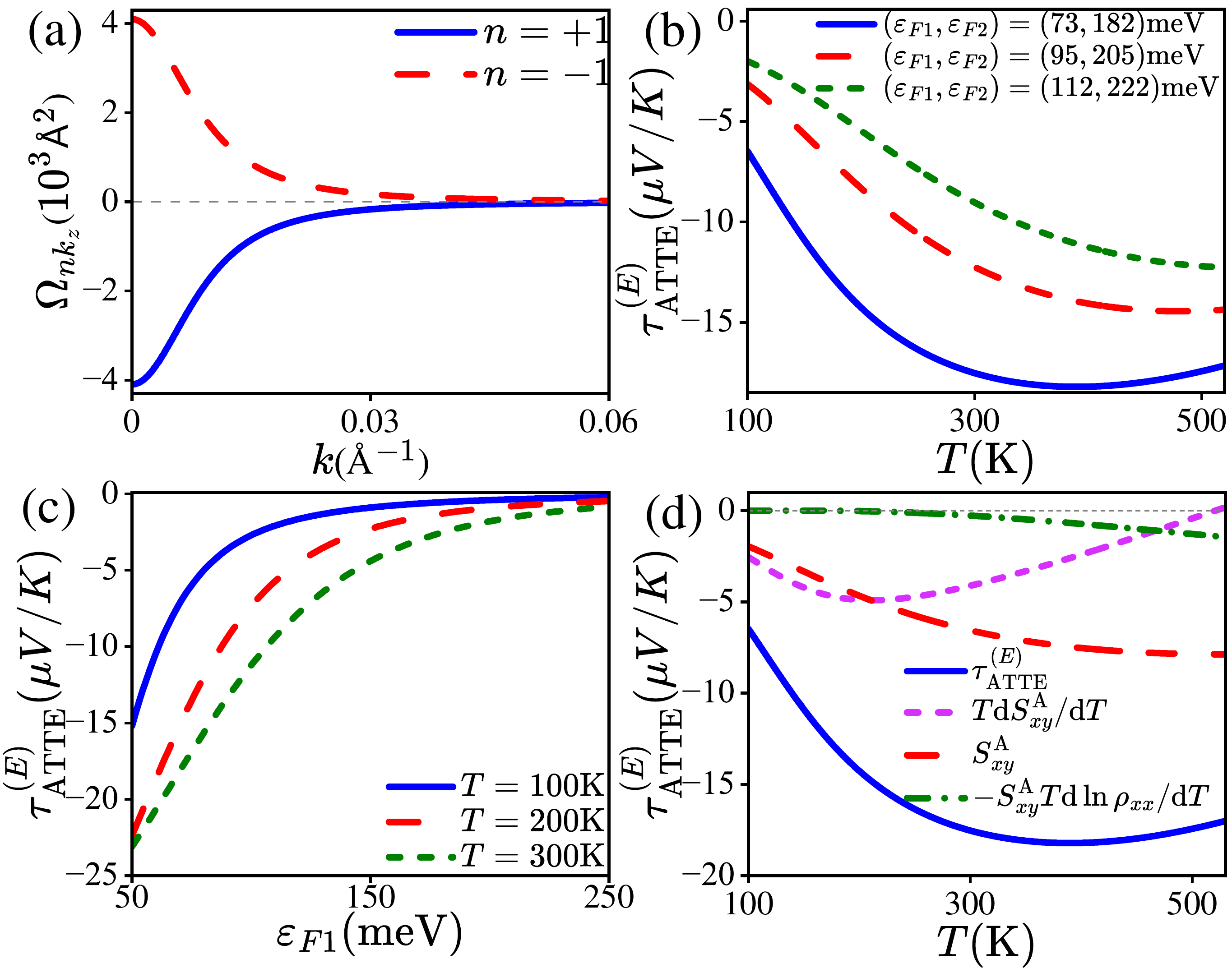}
	\caption{ (a)  Radial dependence  of BC  $\Omega_{\pm,k_z}$  in momentum space.(b) $\CVA\tau^{(E)}_{\text{ATTE}}\CIVA$ vs temperature $T$  for representative two-cone Fermi energy pairs ($\varepsilon_{F1}, \varepsilon_{F2}$).   (c)   $\CVA\tau^{(E)}_{\text{ATTE}}\CIVA$ vs Fermi level  $\varepsilon_{F1}$. (d) The \CVA ATTE \CIVA coefficient, $T\frac{\mathrm{d} S_{xy}^{\text A}}{\mathrm{d} T}$, $S_{xy}^{\text A}$ and $-S_{xy}^{\text A}T\frac{\text d\ln\rho_{xx}}{\text dT}$ vs temperature $T$  for ($\varepsilon_{F1}, \varepsilon_{F2}$) =(73, 182) meV. The parameters are set as $a = 5.338$ \AA, $\Delta= 16$  $\text{meV}$, $v=$ 1.448 $\text{eV\AA} $ and $\tau_s$ = 21.56 $\text{fs}$ \cite{Ye2018,Ye2019,Li2026,Papaj2021}.
	}
	\label{Fig.2}
\end{figure}

 \textit{\CVA ATTE \CIVA in Dirac semimetal.-}
To gain a quantitative understanding of \CVA ATTE\CIVA, we perform model calculations for topological layered kagome ferromagnet Fe$_3$Sn$_2$ \cite{Ye2018,Ye2019,Li2023,Tao2023,Khan2024,Li2019,Wang2020,Lyalin2021} with time-reversal symmetry breaking.
Fe$_3$Sn$_2$ \CVB contains a pair of quasi-two-dimensional massive Dirac cones near the Fermi level \CIVB due to the interplay of spin-orbit coupling and ferromagnetism, and exhibits pronounced responses in both the Berry curvature-induced ANE and AHE \cite{Li2023,Ye2018}.
The Hamiltonian \cite{Papaj2021,Ye2018,Ye2019} reads
\begin{equation}\label{eq:H(k)_Fe3Sn2}
	H=v( sk_{x} \sigma _{x} +k_{y} \sigma _{y}) +\Delta \sigma _{z},
\end{equation}
where $v$ is Fermi velocity,  $s=\pm1$ stands for chirality, $\Delta$ is the gap of Dirac cone, $k_i$ and $\sigma_{i}$ $(i \in \{x,y,z\})$  denote the wave vector and Pauli matrix, respectively.
The band energy is $\varepsilon _{nk } =n \sqrt{\Delta ^{2} +v^{2} k^{2}}$ ($n=\pm$ for conduction (valence) band) and the Berry curvature (BC) is $\Omega _{nk_z} =- \frac{nsv^{2} \Delta }{2\left( \Delta ^{2} +v^{2} k^{2}\right)^{3/2}}$.
\CVB Fig. \ref{Fig.2}(a) shows its radial dependence \(\Omega_{n k_z}\), which is sharply peaked near the gapped Dirac point and decays rapidly at larger \(k\). \CIVB
%Fig. \ref{Fig.2}(a) shows $\Omega_{+,k_z}$ mapped to $(k_x,k_y)$ plane, which exhibits a pronounced peak near the Brillouin zone center and decays rapidly to nearly zero away from this region.
%%%%%%%%%%%%%%%%%%%%%%%%%%%%%%%%%%%%%%%%%%%%%%%%%%%%%%%%%%%%%%%%%%%%%%%%%%%%
%%%%%%%%%%%%%%%%%%%%%%%%%%%%%%%%%%%%%%%%%%%%%%%%%%%%%%%%%%%%%%%%%%%%%%%%%%%%%%%%%%%%%%%%%%%%%%%%%%%%%%%%%%%%%%%%%%%%%%%%%%%%%%%%%%%%
%%%%%%%%%%%%%%%%%%%%%%%%%%%%%%%%%%%%%%%%%%%%%%%%%%%%%%%%%%%%%%%%%%%%%%%%%%%%%%%%%%%%%%%%%%%%%%%%%%%%%%%%%%%%%%%%%%%%%%%%%%%%%%%%%%%

To derive the \CVA ATTE \CIVA of Fe$_3$Sn$_2$, we proceed by computing the \CVA transport coefficients \(\sigma_{xx}\),  \(\alpha_{xx}\), \(\sigma^{\text A}_{xy}\), and \(\alpha^{\text A}_{xy}\)  \CIVA within Boltzmann transport equation \cite{sm}\CIVA.
$\sigma _{xx}$ and $\alpha _{xx}$ are related to the relaxation time $\tau _{s} $= 21.56 $\text{fs} $ \cite{Khan2024,Madsen2006,Li2023,Katsura2018,Aziz2016,Katsura2018,Faghaninia2015,Kinaci2015}, denoting the %characteristic 
effective scattering time in the material \cite{sm}. \CVA The $\sigma _{xx}$ and $\alpha _{xx}$ are controlled by the Fermi-window factor $\partial f_{nk}^{0}/\partial \varepsilon _{nk}$. The $\sigma^{\text A}_{xy}$ is a BC-weighted Fermi sea response, whereas the $\alpha^{\text A}_{xy}$ is weighted by the entropy density \cite{Yu2015,Zhang2024} $S_{nk}= -f_{nk}^{0}\ln f_{nk}^{0} -( 1-f_{nk}^{0})\ln( 1-f_{nk}^{0})$ and is therefore concentrated within a few \(k_BT\) of the Fermi level. \CIVA
%
%The $\sigma^{\text A}_{xy}$ and $\alpha^{\text A}_{xy}$ are composed by BC and the entropy density \cite{Yu2015,Zhang2024} $S_{nk}= -f_{nk}^{0}\ln f_{nk}^{0} -( 1-f_{nk}^{0})\ln( 1-f_{nk}^{0})$.
%
The equilibrium Fermi-Dirac distribution reads $f_{nk}^{0} =(1+e^{{(\varepsilon _{nk} -\varepsilon _{F})}/{k_{B} T}})^{-1}$. %, whose energy derivative determines the thermal broadening of the transport window, satisfying $\frac{\partial f_{nk}^{0}}{\partial \varepsilon _{nk}}  =-\frac{1}{k_{B} T} f_{nk}^{0} ( 1-f_{nk}^{0} )$.
The $S_{nk}$ and $\partial f_{nk}^{0}/\partial \varepsilon _{nk}$ confine the dominant contribution \CVA to states within a few \(k_BT\) of the Fermi energy \(\varepsilon_F\). \CIVA
%a narrow energy range of approximately a few $k_B T$ around Fermi energy $\varepsilon_F$ \cite{Yu2015,Zhang2024}.
%
%
%\CV
$k_c=(2\sqrt{\pi})/(3^{\frac{1}{4}}a)$ \cite{sm} is a wave vector cutoff from the Debye model \cite{Kittel2013, Yu2015} approximation.
%

%%%%%%%%%%%%%%%%%%%%%%%%%%%%%%%%%%%%%%%%%%%%%%%%%%%%%%%%%%%%%%%%%%%%%%%%%%%%%%%%%%%%%%%%%%%%%%%%%%%%%%%%%%%%%%%%%%%%%%%%%%%%%%%%%%%%%%%%%%%

%By evaluating the integrals in
%From Eqs. (\ref{eq:sigma_xx_Fe3Sn2})-(\ref{eq:alpha_xy_Fe3Sn2}), we obtain the temperature and Fermi energy dependence of $\CVA\tau_{\text{ATTE}}\CIVA$.
%
\CVB Fe$_3$Sn$_2$ hosts two energy-split massive Dirac cones near the Fermi level. We retain the local single cone Hamiltonian Eq. (\ref{eq:H(k)_Fe3Sn2}) and incorporate the energy splitting through the cone-resolved Fermi energies ($\varepsilon_{F1}$,$\varepsilon_{F2}$). \CVA The transport coefficients \CIVA are evaluated for each cone and summed over both cones, from which $\CVA\tau^{(j)}_{\text{ATTE}}$, $S_{xy}^{\text A}$ and $\CVA\tau^{(E)}_{\text{ATTE}}$ \CIVA are obtained using Eqs. (\ref{eq:tau_ATTE})-(\ref{eq:tau_ATTE_E}), respectively. Fig. \ref{Fig.2}(b) shows $\CVA\tau^{(E)}_{\text{ATTE}}(T)\CIVA$ for representative Fermi-energy pairs ($\varepsilon_{F1}$,$\varepsilon_{F2}$)=(73, 182), (95, 205) and (112, 222) meV \cite{Ye2018,Ye2019}. \CIVB
%
%The calculation assumes that  $\varepsilon_F$ lies in the conduction band, with its value set to be experimental values of 118 meV, 125 meV, and 207 meV \cite{Li2023,Ye2018}, and we consider a single Dirac cone. %\CIV
%
%After performing the integrals in Eqs. (\ref{eq:sigma_xx_Fe3Sn2})-(\ref{eq:alpha_xy_Fe3Sn2}), %we obtain the temperature and Fermi energy dependence of $\CVA\tau_{\text{ATTE}}\CIVA$.
%
%$\CVA\tau_{\text{ATTE}}\CIVA$ vs $T$ for different $\varepsilon_F$ are derived in Fig. \ref{Fig.2}(b). %, we present the influence of ATE in a range of 100 K to 500 K.
%
The magnitude of $\CVA\tau^{(E)}_{\text{ATTE}}\CIVA$ reaches \CVB $-18.2$ $\mu$V/K around 389 K at ($\varepsilon_{F1}$,$\varepsilon_{F2}$)=(73, 182) meV, and the negative sign of $\CVA\tau_{\text{ATTE}}\CIVA$ corresponds to heat absorption under our sign convention, for the fixed geometry in Fig. \ref{Fig.1}(i). \CIVB
%
%and the negative sign of $\CVA\tau_{\text{ATTE}}\CIVA$ signifies a transverse cooling effect within the material.
%
Notably, the result reveals that the  \CVA magnitude of the reversible ATTE contribution \CIVA is enhanced as the Fermi energy decreases, i.e., the closer $\varepsilon_F$ to the Dirac point, the more pronounced the \CVA ATTE \CIVA occurs.

%%%%%%%%%%%%%%%%%%%%%%%%%%%%%%%%%%%%%%%%%%%%%%%%%%%%%%%%%%%%%%%%%%%%%%%%%%%%%%%%%%%%%%%%%%%%%%%%%%%%%%%%%%%%%%%%%%%%%%%%%%%%%%%%%%%%
%\begin{figure}[tb]
%	\centering
%	\includegraphics[width=1\linewidth]{Fig2.pdf}
%	\caption{ (a) Contour plots of BC $\Omega_{+,k_z}$ in momentum space for the conduction band.(b) ATE coefficient $\CVA\tau_{\text{ATTE}}\CIVA$ vs temperature $T$.   (c) ATE coefficient $\CVA\tau_{\text{ATTE}}\CIVA$ vs Fermi level $\varepsilon_{F}$. (d) The ATE coefficient, $T\frac{\mathrm{d} S_{xy}^{A}}{\mathrm{d} T}$ and $S_{xy}^{A}$ vs temperature $T$ at $\varepsilon_{F}=73$ meV. The parameters are set as $a = 5.338$ \AA, $\Delta= 16$  $\text{meV}$, $v=$ 1.448 $\text{eV \AA}$ and $\tau_s$ = 35.6 $\text{fs}$ \cite{Ye2018,Ye2019,Li2024,Papaj2021}.}
%	\label{Fig.2}
%\end{figure}
%%%%%%%%%%%%%%%%%%%%%%%%%%%%%%%%%%%%%%%%%%%%%%%%%%%%%%%%%%%%%%%%%%%%%%%%%%%%%%%%%%%%%%%%%%%%%%%%%%%%%%%%%%%%%%%%%%%%%%%%%%%%%%%%%%%

Fig. \ref{Fig.2}(c) shows the evolution of $\CVA\tau^{(E)}_{\text{ATTE}}\CIVA$ as a function of  \CVB $\varepsilon_{F1}$, with $\varepsilon_{F2}=\varepsilon_{F1}+\Delta E$ and $\Delta E \approx 110$ meV \cite{Ye2019}. \CIVB It is seen that $\CVA\tau^{(E)}_{\text{ATTE}}\CIVA$ is monotonically increased as $\varepsilon_F$ shifts closer to the conduction band edge.
%
%This enhancement is attributable to the increased BC when approaching band edge.
This trend primarily reflects the enhancement of BC-driven anomalous transport near the band edge, which strengthens the ANC-related terms in Eq. (\ref{eq:tau_ATTE_E}).
Consequently, the local  heat absorption capacity, which is directly related to $\CVA\tau^{(E)}_{\text{ATTE}}\CIVA$, is optimized near the band edge, confirming that tuning the Fermi level towards the band edge is critical for optimizing the transverse heat management.
In Fig. \ref{Fig.2}(d), we show how $T \mathrm{d} S_{xy}^{\text{A}}/\mathrm{d} T$,  $S_{xy}^{\text A}$, $-S_{xy}^{\text A}T\text d\ln\rho_{xx}/\text dT$ and $\CVA\tau^{(E)}_{\text{ATTE}}\CIVA$ vary with temperature.
%Additionally, Eq. (\ref{eq:tau_ATTE}) delineates that the ATTE comprises two distinct components $T\frac{\mathrm{d} S_{xy}^{A}}{\mathrm{d} T}$ and $2S_{xy}^{A}$. We provide the detailed visualization of the competitive interplay between these terms in Fig. \ref{Fig.2}(d).
\CVB The $T \mathrm{d} S_{xy}^{\text A}/\mathrm{d} T$ term, $S_{xy}^{\text A}$ term and $-S_{xy}^{\text A}T\text d\ln\rho_{xx}/\text dT$ term may either reinforce or partially compensate each other, depending on whether $T \mathrm{d} S_{xy}^{\text A}/\mathrm{d} T$ changes its sign. \CIVB %This illustrates that $\CVA\tau_{\text{ATTE}}\CIVA$ is controlled not only by the magnitude of $S_{xy}^A$, but also by its temperature variation.
For $(\varepsilon_{F1}, \varepsilon_{F2})=(73, 182)$ meV \cite{Ye2019} and $T>521 $K,
%the two constituent components of ATC take opposite signs:
\CVA $T \mathrm{d} S_{xy}^{\text A}/\mathrm{d} T$ becomes positive while $2S_{xy}^{\text A}$ and $-S_{xy}^{\text A}T\text d\ln\rho_{xx}/\text dT$ remain negative. \CIVA
%
%Their opposing contributions partially cancel, reflecting their competition.
%
%When $\mathrm{d} S_{xy}^{A}/\mathrm{d} T=0$,  $\tau_{\mathrm{ATE}}=2S^{A}_{xy}$, i.e., the ATE is exactly twice the ANE.
%
%If $\mathrm{d} S_{xy}^{A}/\mathrm{d} T>0$, $\tau_{\mathrm{ATE}}>2S^{\text A}_{xy}$, indicating an enhancement.
%
%Conversely, if $\mathrm{d} S_{xy}^{A}/\mathrm{d} T<0$, $\tau_{\mathrm{ATE}}<2S^{A}_{xy}$  (assuming $S^{\text A}_{xy}>0$), thus the ATE may even reverse sign compared to $S^{\text A}_{xy}$, reflecting that the ATE is jointly determined by the two terms.
%
  % In main text, concise one-two sentence summary
We note that the calculated $\sigma^{\text A}_{xy}$, $\alpha^{\text A}_{xy}$, and $S^{\text A}_{xy}$ are compared with available experimental data from Ref. \cite{Ye2018,Li2023,Zhang2021}, showing qualitative consistency.
%Detailed discussion is provided in the Supplementary Material \cite{sm}.

%%%%%%%%%%%%%%%%%%%%%%%%%%%%%%%%%%%%%%%%%%%%%%%%%%%%%%%%%%%%%%%%%%%%%%%
%
%Furthermore, we can obtain analytical expressions for ANC and ATC for $T\to0$ limit
 Furthermore, the low-temperature behavior of the \CVA ATTE \CIVA can be obtained analytically \cite{sm}. For the model in Eq. (\ref{eq:H(k)_Fe3Sn2}), via the Mott relation and Sommerfeld expansion, \CVA $S^{\text A}_{xy} \propto T$, while $T \text{d}\sigma_{xx}/\text{d}T\to 0$, %Hence both $\tau^{(j)}_{\text{ATTE}}/S_{xy}^{\text A}$ and $\tau^{(E)}_{\text{ATTE}}/S_{xy}^{\text A}$ converges to 3.
  yielding \cite{sm}\CIVA
  %Therefore, Eq. (\ref{eq:tau_ATTE}) shows that the \CVA ATTC\CIVA/ANC ratio  converges to 3 in the low-temperature limit.
%\CV
\begin{equation}
	\begin{aligned}
		%S_{xy}^{\text A}(T\to0)&  =-\frac{h\pi s\Delta \varepsilon _{F} k_{B} T}{3\tau_s \left( \Delta ^{2} -\varepsilon _{F}^{2}\right)^{2}}\frac{k_{B}}{e},\\
	\CVA	\tau^{(j/E)}_{\text{ATTE}}  \CIVA & =3S_{xy}^{\text A} %= -\frac{h\pi s\Delta \varepsilon _{F} k_{B} T}{\tau_s \left( \Delta ^{2} -\varepsilon _{F}^{2}\right)^{2}}\frac{k_{B}}{e}\\
		=\frac{-4\pi^3\tau_shs\Delta\varepsilon_Fk_BT}{\left({h}^2\Delta^2 + 4\pi^2\left(\Delta^2-\varepsilon_F^2\right)^2\tau_s^2\right)}\frac{k_B}{e},
	\end{aligned}
\label{tauSxy}
\end{equation}
 where the prefactor is explicitly model-dependent.
%which shows that both ANE and ATE depend linearly on $T$, and in the limit $T \rightarrow 0$K  one obtains $\CVA\tau_{\text{ATTE}}\CIVA=3S^{\text{A}}_{xy}$. Therefore, the ATE is generally three times stronger than the ANE at low temperature.

%%%%%%%%%%%%%%%%%%%%%%%%%%%%%%%%%%%%%%%%%%%%%%%%%%%%%%%%%%%%%%%%%%%%%%%%%%%%%%%%%%%%%%%%%%%%%%%%%%%%%%%%%%%%%%%%%%%%%%%%%
\begin{figure}[tb]
	\centering
	\includegraphics[width=1\linewidth]{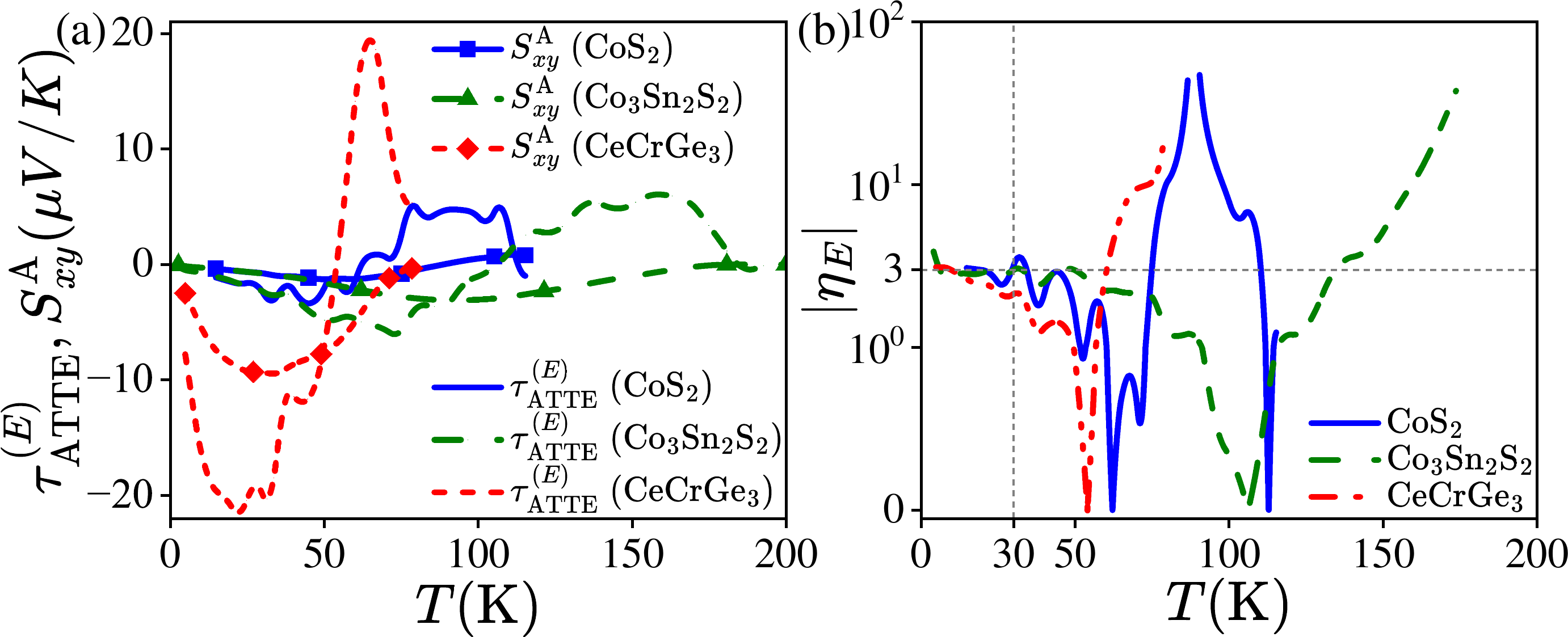}
	\caption{(a) Temperature dependence of the ANC $S^{\text A}_{xy}$ (line-symbol type, taken from experiments)  and the calculated $\CVA\tau^{(E)}_{\mathrm{ATTE}}\CIVA$ (line type) for CoS$_2$, Co$_3$Sn$_2$S$_2$, and CeCrGe$_3$, where $\CVA\tau^{(E)}_{\mathrm{ATTE}}\CIVA$ is evaluated from the experimental $S^{\text A}_{xy}$ and $\rho_{xx}$ using Eq.(\ref{eq:tau_ATTE_E}). (b) The ratio $\eta_E=\CVA\tau^{(E)}_{\text{ATTE}}\CIVA/S_{xy}^{\text{A}}$ vs $T$.  Note: \(|\eta_E|\) on a logarithmic scale over a wider temperature range $T\in[0,200]$K is shown.  }
	\label{Fig.3}
\end{figure}
%%%%%%%%%%%%%%%%%%%%%%%%%%%%%%%%%%%%%%%%%%%%%%%%%%%%%%%%%%%%%%%%%%%%%%%%%%%%%%%%%%%%%%%%%%%%%%%%%%%%%%%%%%%%%%%%%%%%%%%%%

 \textit{\CVA ATTE \CIVA coefficients for several realistic materials.-}
Eqs. (\ref{eq:tau_ATTE}) and (\ref{eq:tau_ATTE_E}) introduce a novel physical quantity that not only enriches our understanding in physics but also holds broad applicability across diverse materials.
	To demonstrate this, we utilize experimentally measured ANC $S^{\text{A}}_{xy}$ \CVA and  longitudinal  resistivity $\rho_{xx}$ data \CIVA from literature \cite{Zhang2025,Li2026,Dalui2025,Guin2019} to evaluate the \CVA $\tau^{(E)}_{\text{ATTE}}$ \CIVA in several representative materials.
Fig. \ref{Fig.3}(a) illustrates the calculated \CVA $\tau^{(E)}_{\text{ATTE}}$ \CIVA for CoS$_2$, Co$_3$Sn$_2$S$_2$, and CeCrGe$_3$, alongside their corresponding measured $S^{\text{A}}_{xy}$ values \cite{Zhang2025,Li2026,Dalui2025,Guin2019}. Notably, CeCrGe$_3$ exhibits \CVA $\tau^{(E)}_{\text{ATTE}}$ \CIVA reaching up to -20 $\mu V/K$, \CVA identifying it as a promising candidate for a direct ATTE
heat-source measurement. \CIVA

% suggesting its potential as an excellent low-temperature refrigeration material.
%%%%%%%%%%%%%%%%%%%%%%%%%%%%%%%%%%%%%%%%%%%%%%%%%%%%%%%%%%%%%%%%%%%%%%%%%%%%%%%%%%%%%%%%%%%%%%%%%%%%%%%%%%%%%%%%%%%%%%%%%%%%%%%%%%%%%%%%%%%%%

%To compare the \CVA ATTE \CIVA and ANE,
 Fig. \ref{Fig.3}(b) exhibits the variation of the dimensionless ratio $\eta_E=\CVA\tau^{(E)}_{\text{ATTE}}\CIVA/S_{xy}^{\text{A}}$ with $T$ for three materials. \CVA When \(S_{xy}^{\rm A}\) is close to zero, \(\eta_E\) becomes sensitive to its small denominator. We therefore exclude these points when interpreting the ratio. \CIVA
At low temperatures ($T < 30$ K), $|\eta_E|$ hovers around 3, \CVA qualitatively consistent with the predicted low-temperature trend \(\eta_E\to3\). \CIVA
%which is consistent with \CVB the result \CIVB $\CVA\tau^{(E)}_{\text{ATTE}}\CIVA=3S^{\text{A}}_{xy}$ in the zero-temperature limit.
%
As $T>30$ K, $\eta_E$ begins to systematically deviate from \CVA 3 as both $S_{xy}^{\text A}(T)$ and $\sigma_{xx}(T)$ evolve. \CIVA
%
%Notably, CeCrGe$_3$ exhibits both the largest $|S^{\text{A}}_{xy}|$ and $\CVA|\tau^{(E)}_{\text{ATTE}}|\CIVA$ among these three materials.
%
\CVA  The magnitude $|\eta_E|$ shows a nonmonotonic, material-dependent temperature dependence. Near liquid-nitrogen temperatures, $|\eta_E|$ for CeCrGe$_3$ exceeds 10, while CoS$_2$ exhibits a pronounced peak near 90 K [Fig.~\ref{Fig.3}(b)]. Over the temperature range shared by the three datasets, CeCrGe$_3$ has the largest $|S_{xy}^{\rm A}|$ and $|\tau_{\rm ATTE}^{(E)}|$, making it a promising candidate for direct measurement of the local ATTE heat source.  \CIVA

%With increasing $T$, $\eta_E$ generally increases. For instance, in the liquid nitrogen temperature range, $\eta_E$ for CeCrGe$_3$ can exceed 10, indicating that the \CVA $\tau^{(E)}_{\text{ATTE}}$ \CIVA is approximately 10 times larger than the ANC. At even higher temperatures, as shown in the logarithmic inset of Fig. \ref{Fig.3}(b), \(\eta_E\) for CoS\(_2\) can increase to values of order \(10^2\). The sizable values of both \(\left|S_{xy}^{A}\right|\) and \(\left|\tau_{\rm ATTE}^{(E)}\right|\) make the material favorable for a direct measurement of the local ATTE heat source.
%
%\CVA This significant enhancement highlights its promising potential for solid-state cooling applications.  $\eta$ serves primarily as a fundamental dimensionless transport diagnostic rather than a direct figure of merit for cooling, and this motivates a direct, quantitative evaluation of its actual macroscopic cooling performance, which we address next. \CIVA%At even higher $T$ (e.g., above 80 K, not shown), $\eta$ can be several hundreds.
%
%%%%%%%%%%%%%%%%%%%%%%%%%%%%%%%%%%%%%%%%%%%%%%%%%%%%%%%%%%%%%%%%%%%%%%%%%%%%%%%%%%%%%%%%%%%%%%%%%%%%%%%%%%%%%%%%%%%%%%%%%

 \textit{The Maximization of the Heat current $j^{\text{max}}_q$.-}
 \CVA To assess the practical cooling advantage of the ATTE, it is necessary to go beyond $\eta_E$ and evaluate a quantitative thermoelectric cooling capacity.  \CIVA For the fixed convention used in Fig. \ref{Fig.1}(i),  when $\CVA\tau^{(E)}_{\text{ATTE}}\CIVA < 0$, heat is absorbed within the material, providing a pathway for cooling.
 We consider the maximum cold side heat current density $j^{\text{max}}_{q,y}$ \cite{Su2018,Chen2022,Jeong2021} as a simple analytical measure.
\CVA  $j^{\text{max}}_{q,y}$ represents the maximum heat current density where the cold end is allowed to pass through for a device, which is determined by materials and geometry.\CIVA

In \CVA a \CIVA steady state, we impose the boundary condition of $T(y=0)=T_C$, $T(y=L)=T_H$ ($L$ is length), \CVA with $\Delta T=T_H-T_C>0$.
%For continuous equipotential electrodes, $E_x$ is uniform, whereas the external circuit fixes $\bar j_{c,x}=L^{-1}\int_0^L j_{c,x}(y)dy$. Optimizing this average current, we obtain the achievable maximization of the heat current density $j_{q,y}^{C,\text{max}}$ through the cold side gives \cite{sm}
For equipotential electrodes, $E_x$ is uniform along the transverse direction, while the external circuit fixes the average current density, $\bar j_{c,x}=L^{-1}\int_0^L j_{c,x}(y)dy$. Maximizing the heat current density at the cold side with respect to this average current yields the optimum $\bar j^{\rm opt}_{c,x}$ and hence the maximum achievable heat current density at the cold side $j_{q,y}^{C,\rm max}$ \cite{sm}:
	\begin{equation}
		j_{q,y}^{C,\max}
		=\frac{\kappa_C'}{L}
		\left[-\Delta T+\frac{z_CT_C^2}{2}
		\left(1+\frac{\beta_{j,C}}{2}\right)^2\right].
		\label{eq:jmax_dimensionless}
	\end{equation}
%	 The first term is conductive backflow, while the second is the optimized active transverse pumping contribution. Eq. (\ref{eq:jmax_dimensionless}) is governed by $\kappa_C'=\kappa_0'(T_C)$, the thermoelectric figure of merit $	z_CT_C={(S_{C,xy}^{A})^2}T_C/{(\rho_C\kappa_C')}$, 	 and the dimensionless quantities $\eta_{j,C}={\tau_{{\rm ATTE},C}^{(j)}}/{S_{C,xy}^{\text A}}$, $\beta_{j,C}=\eta_{j,C}{\Delta T}/{T_C}$ at the cold side.
 The first term represents conductive backflow, whereas the second is the optimized active transverse pumping contribution. The maximum heat current density in Eq. (\ref{eq:jmax_dimensionless}) is governed by the cold-side thermoelectric figure of merit $z_CT_C = (S^{\rm A}_{C,xy})^2 T_C / (\rho_C\kappa_C')$, where $\kappa_C'=\kappa_0'(T_C)$, together with two dimensionless quantities evaluated at the cold side,
	$\eta_{j,C}={\tau_{{\rm ATTE},C}^{(j)}}/{S_{C,xy}^{\text A}}$ and $\beta_{j,C}=\eta_{j,C}{\Delta T}/{T_C}$.
	
Relative to the same material with the ATTE term omitted, this active contribution is enhanced by the factor \cite{sm}
%Relative to the zero-\(\Delta T\) anomalous Ettingshausen pumping reference, the finite $\nabla T$ correction is quantified by
\begin{equation}
		\mathcal{G}_C
		=
		\left(1+{\beta_{j,C}}/{2}\right)^2.\label{eq:G_revised}
\end{equation}
% For CeCrGe$_3$ \cite{Li2026}, at $T_C=12.48$K and $\Delta T=1$K and$	\mathcal G_C\simeq1.23$.Thus, for the configuration in which the ATTE assists heat extraction at the cold side, the optimized reversible heat pumping contribution is enhanced by about \(23\%\). This enhancement refers to the reversible pumping contribution rather than the net cooling flux, which also depends on \(\rho_C\), \(\kappa_C'\), device geometry, thermal contact resistances, and conductive heat backflow.	
For CeCrGe$_3$ \cite{Li2026} at $T_C = 12.48$ K with $\Delta T=1$K, Eq. (\ref{eq:G_revised}) gives $\mathcal G_C\simeq1.23$. Thus, in the configuration where the ATTE assists heat extraction at the cold side, the optimized reversible pumping contribution is enhanced by about 23$\%$. We note that this enhancement applies to the reversible pumping contribution, not to the net cooling flux, which depends on \(\rho_C\), \(\kappa_C'\), device geometry, thermal contact resistances, and conductive heat backflow.
	 	 \CIVA

%
%Optimizing with respect to the applied charge current \CVB$dj^C_{q,y}/dj_{c,x}=0$ \CIVB yields Accordingly, we yield
%\begin{equation}
%	\begin{aligned}
%		j_{q,x}^{\text{max}} =-\frac{\kappa '_{xx} T_{C}}{L}\left[\frac{3z_{C} T_{C}( 2+\beta \theta _{th})^{2}}{8( \beta z_{C}T_{C}\eta '_{C}  -3) \theta _{th}} +\frac{\Delta %T}{T_{C}} \theta _{th}\right].
%	\end{aligned}
%\label{eq:Delta T_{max}}
%\end{equation}

%
Consequently, Eq. (\ref{eq:jmax_dimensionless}) \CVB makes explicit that the \CVA ATTE \CIVA modifies the maximum cold side heat current density $j_{q,y}^{\text{max}}$ through $\eta_{j,C}$ and $\beta_{j,C}$, \CIVB and serves as a quantitative metric for assessing and comparing the refrigeration capabilities of diverse materials. This implies that the capacity of thermoelectric coolers \CVA has the fundamental potential to \CIVA be enhanced through adjusting the material's figure of merit, and by strategically harnessing the ATTE. \CVA %Ultimately, dedicated experimental measurements of the local reversible heat absorption/release under the mutually transverse current and temperature-gradient geometry will be essential to confirm these predictions in practical applications.
Looking ahead, direct measurements of the local reversible heat absorption and release in the mutually transverse current and temperature-gradient geometry will be important to experimentally confirm these predictions, and would firmly establish the ATTE as a member of the transverse thermoelectric family. \CIVA
%

%%%%%%%%%%%%%%%%%%%%%%%%%%%%%%%%%%%%%%%%%%%%%%%%%%%%%%%%%%%%%%%%%%%%%%%%%%%%%%%%%%%%%%%%%%%%%%%%%%%%%%%%%%%%%%%%%%%%%%%%%

%\section{Conclusions and Discussions}
%\label{section6}
%In summary, we propose the anomalous Thomson effect and consider the Fe$_3$Sn$_2$ magnetic materials. We find that the ATE cooling effect performs better around 400K at $\varepsilon_{F} = 118 \text{meV}$. Moreover, when the Fermi level is adjusted to the bottom of the conduction band, the ATTE coefficient becomes larger. Although we tried our best to obtain the ATTE coefficient in the calculation, we only consider the intrinsic contribution of the anomalous effect, which also includes side jump and skew scattering mechanisms \cite{Xiao2010,Nagaosa2010}, and we did not take into account. Furthermore, in the given ATTE results, the influence of relaxation time $\tau_s$ is not precisely considered, and it is highly related to different samples. Therefore, the fact that our calculation results have some reference value does not mean that such a conclusion can be drawn for any sample.

%\begin{widetext}
%\appendix
%\section{ATTE}
%\end{widetext}

%\begin{acknowledgments}
%	
 \textit{Acknowledgments.-}
This work is supported by the National Key R\&D Program of China (Grant No. 2024YFA1409200, No. 2022YFA1402802), CAS Project for Young Scientists in Basic Research Grant No. YSBR-057.  G.S. was supported in part by the Quantum Science and Technology-National Science and Technology Major Project under Grant No. 2024ZD0300500, NSFC Nos. 12534009 and 12447101, the Strategic Priority Research Program of CAS (Grant No. XDB1270000) and the CAS Superconducting Research Project under Grant No. SCZX-0101.  ZFZ was supported by the Postdoctoral Fellowship Program of CPSF under Grant No. GZB20250793. %\CIV
%\end{acknowledgments}

%	
%
%\end{widetext}

%\bibliography{thomson_main_ref}
%apsrev4-2.bst 2019-01-14 (MD) hand-edited version of apsrev4-1.bst
%Control: key (0)
%Control: author (8) initials jnrlst
%Control: editor formatted (1) identically to author
%Control: production of article title (0) allowed
%Control: page (0) single
%Control: year (1) truncated
%Control: production of eprint (0) enabled
%

\end{document}